\documentstyle[12pt,aasms4]{article}
\lefthead{Jones et al.}
\righthead{Interstellar Scattering of PKS~1830-211}
\begin{document}
\title{Interstellar Broadening of Images in the\\ 
Gravitational Lens PKS~1830-211}   

\author{D.L.~Jones, R.A.~Preston, D.W.~Murphy}
\affil{Jet Propulsion Laboratory, California Institute of Technology,
Pasadena, CA. 91109}

\author{D.L.~Jauncey, J.E.~Reynolds, A.K.~Tzioumis, E.A.~King} 
\affil{Australia Telescope National Facility, CSIRO, Epping, NSW,
Australia} 

\author{P.M.~McCulloch, J.E.J.~Lovell, M.E.~Costa}   
\affil{Dept.~of Physics, University of Tasmania, Hobart,
Tasmania, Australia}

\and

\author{T.D.~van Ommen}
\affil{Antarctic CRC, University of Tasmania,
Hobart, Tasmania, Australia}

\begin{abstract}
The remarkably strong radio gravitational lens PKS~1830-211 
consists of a one arcsecond diameter Einstein ring with two bright 
compact (milliarcsecond) components located on opposite sides of 
the ring.  We have obtained 22 GHz VLBA data on this source to  
determine the intrinsic angular sizes of the compact components.  
Previous VLBI observations at lower frequencies indicate that the 
brightness temperatures of these components are significantly  
lower than $10^{10}$ K (Jauncey, et al. \markcite{1} 1991), less than  
is typical for compact synchrotron radio sources and less than 
is implied by the short timescales of flux density variations.  
A possible explanation  
is that interstellar scattering is broadening the apparent angular 
size of the source and thereby reducing the observed brightness 
temperature.  Our VLBA data support this hypothesis.  
At 22 GHz the measured brightness temperature is at least $10^{11}$
K, and the deconvolved size of the core in the southwest compact 
component is proportional to $\nu^{-2}$ between 1.7 and 22 GHz. 
VLBI observations at still higher frequencies should be
unaffected by interstellar scattering.  
\end{abstract}


\section{Introduction}

The Einstein ring gravitational lens PKS~1830-211 is one of  
the strongest compact radio sources in the sky, nearly two  
orders of magnitude brighter at radio frequencies than any   
other known gravitational lens (Rao and Subrahmanyan \markcite{2} 
1988; Jauncey, et al. \markcite{1} 1991).  The source consists of 
a pair of bright compact (mas) components on opposite sides of a 1 
arcsecond diameter ring of emission.  Both compact components have 
been imaged with VLBI at 1.7 GHz (Jones, unpublished), 2.3 GHz  
(Jauncey, et al. \markcite{1} 1991; King, et al. \markcite{3} 1993),
5 GHz (Jones, et al. \markcite{4} 1993; Jones \markcite{5} 1994), 
and 22 GHz (Jones, et al. \markcite{6}   
1995 and this paper).  Variations in the morphology 
of both compact components have been seen between two VLBI epochs 
at 5 GHz (Jones \markcite{5} 1994).  The extragalactic nature of 
PKS~1830-211 was confirmed by galactic HI emission and absorption 
measurements (Subrahmanyan, Kesteven, and te Lintel Hekkert 
\markcite{7} 1992).
 
The large radio flux density of PKS~1830-211 has allowed a large 
quantity of high signal-to-noise single-dish and interferometric data to 
be obtained (e.g., Subrahmanyan, et al. \markcite{8} 1990; Jauncey,  
et al. \markcite{9} 1992, \markcite{10} 1993; van Ommen, et al.
\markcite{11} 1995; Lovell, et al. \markcite{12} 1995).   
This wealth of observational information provides strong 
constraints on the geometry of the lens system and may lead to 
an accurate estimate of the large-scale value of ${\rm H}_{\circ}$ 
when combined with redshifts and the differential time delay 
between the two images.  

The over-all morphology of the source has
been successfully modeled as a background core-jet radio source
lensed by a single medium-mass galaxy (Kochanek and Narayan 
\markcite{13} 1992; Nair, Narasimha, and Rao \markcite{14} 1993).  
An optical redshift for the background source is still unavailable 
because the position of PKS~1830-211 is close to the galactic center  
($l=12.\!\!{}^{\circ}2$, $b=-5.\!\!{}^{\circ}7$) 
in a region crowded with foreground stars (Djorgovsky, et al.
\markcite{15} 1992; Jauncey, et al. \markcite{10} 1993).  
However, the redshift of the lensing galaxy has been 
recently determined from radio absorption line observations 
(Wiklind and Combes \markcite{16} 1996).
The time delay between the two compact components has
been estimated as $44 \pm 9$ days by van Ommen, et al.
\markcite{11} (1995) based on multi-epoch VLA images.
Proposed observations with the VSOP spacecraft could improve
the accuracy of the time delay determination.   

One of the aspects of this source which has been difficult to 
explain is the unusually low brightness temperatures of the two
compact components.  The total flux density varies by factors of
more than 2 on time scales of months and years (Lovell, et al.
\markcite{12} 1995), which implies a smaller angular size and higher 
brightness temperature than seen in VLBI images at frequencies 
$\le 5$ GHz.  Since the line of sight to PKS~1830-211 passes close 
to the galactic center, it is plausible that significant interstellar 
scattering (ISS) within our galaxy may occur along this line of 
sight.  Scattering by the interstellar medium in the lensing
galaxy will be far less significant than ISS within our galaxy 
(Walker \markcite{99} 1996).  
The effects of ISS decrease rapidly with frequency, 
and should be significantly smaller at 22 GHz.  

\section{Observations}

We observed PKS~1830-221 with the full VLBA at 22 GHz in May 1994.  
The data were correlated in Socorro twice using phase centers 
corresponding to the locations of the two compact components, 
which we designate the northeast (NE) and southwest (SW) components.  
The observations were inadvertently made in dual-polarization mode,
which resulted in much less suppression of the component far from
the phase center (by decorrelation over the observing bandwidth)
than expected.  As a result, it was necessary to include both 
compact components in the imaging process.
After manual phase calibration, amplitude calibration, and fringe 
fitting in AIPS\footnote{The Astronomical Image Processing System
was developed by the National Radio Astronomy Observatory, which
is operated by Associated Universities, Inc., under a cooperative 
agreement with the National Science Foundation.}, we used the Caltech 
program Difmap (Shepherd, Pearson, and Taylor \markcite{17} 1994) for 
editing, self-calibration, imaging, and deconvolution.  The data were 
coherently averaged to 15 seconds, and the errors were estimated 
from the scatter of the 1-second data points within each averaging
interval.  

Phase-only corrections were applied during each cycle of 
self-calibration, Fourier inversion, and deconvolution  
until the model flux density matched that of the visibility data
on the shortest baselines to within a few percent.  At that point 
time-independent antenna gain corrections were allowed (these 
corrections were $\le 12$\% for all antennas) for several cycles, 
followed by time-dependent gain corrections with decreasing time 
scales until full point-by-point amplitude self-calibration was
applied.  The self-calibration solutions remained very stable from one
cycle to the next.  The final agreement factor (reduced $\chi^{2}$)
between our model and the self-calibrated data was 0.86.

\section{Results}

Figures 1 and 2 show our VLBA images of the NE and SW components 
of PKS~1830-211, respectively.  The NE component appears to have 
a jet extending at least 12 mas towards the NW, and a bright core
that is slightly extended towards the NW as well.  The SW component
appears to be very nearly unresolved.  Recent 15 GHz VLBI 
observations by Garrett, et al. \markcite{18} (1995) confirm 
that the NE component has a well-defined jet as well as a  
core, while the SW component is essentially unresolved
(see also Patnaik and Porcas \markcite{19} 1995). 
The SW component in PKS~1830-211 is the more compact of the two
bright components at all frequencies, and consequently is the best
one to use as a test for ISS broadening.
The 22-GHz image of the SW component shown in Jones, et al.
\markcite{6} (1995) has slightly higher angular resolution but 
significantly lower dynamic range than the image in figure 2.  

\placefigure{fig1}

\placefigure{fig2}

The angular separation between the brightest peaks in figures 1 and
2 is $972 \pm 1$ mas.  VLBI observations of PKS~1830-211 at 5 GHz in 
November 1990 and September 1991 indicated angular separations of 
$975 \pm 2$ and $973 \pm 2$ mas, respectively (Jones, et al.
\markcite{4} 1993).  Thus, we find no evidence for a significant change 
in separation over 3.5 years.  As Williams and Saha \markcite{20} (1995) 
point out, centroid shifts caused by microlensing should not be detectable
in radio images.   

Figure 3 show the deconvolved minor axis 
width of the SW component at four frequencies: 1.7 GHz (unpublished
data from an ad-hoc VLBA experiment in 1990), 2.3 GHz (King \markcite{21}
1995), 4.9 GHz (Jones \markcite{5} 1994), and 22 GHz (this paper).  
Deconvolved sizes are
used to remove (or at least reduce) the bias in observed angular sizes 
which is introduced by the limited range of baseline lengths in each VLBI 
array.  We compare the minor axis sizes because intrinsic source structure
is more likely to contribute to the major axis size.  
The slope of the line fit to the angular size measurements in figure 3 
is $1.96 \pm 0.14$, consistent with the $\lambda^2$ dependence
expected for scattering.  This suggests that angular size measurements
made by VLBI at frequencies $\le 22$ GHz are indeed affected by angular 
broadening due to ISS.  

\placefigure{fig3}

At 22 GHz the deconvolved size of the SW component core is less than $0.6 
\times 0.2$ mas and T$_{\rm b} \ge 10^{11}$ K.  This is much more 
typical of the brightness temperatures seen in other extragalactic 
compact radio sources.  

\section{Discussion}

The amount of ISS expected for a line of sight passing close to the 
center of our galaxy can be estimated from the measured angular 
broadening of isolated H$_{\rm 2}$O maser features in W49(N) and Sgr B2. 
Gwinn, Moran, and Reid \markcite{22} (1988) found minimum angular 
sizes of 0.2 mas in W49 and 0.3 mas in Sgr B2.  This is consistent with
our measured angular size of the SW component core in PKS~1830-211 
at 22 GHz.  Gwinn, Moran, and Reid also found that OH masers in W49  
and Sgr B2 had angular sizes consistent with $\Theta \propto {\lambda}^2$
when compared with H$_{2}$0 maser sizes in those objects.  

It is interesting to compare our measurements with those predicted
by the Blandford and Narayan \markcite{23} (1985) formula for ISS:
$${\Theta}_{\rm ISS} \ \sim \ 2({{C}^{2}}\!\!\!{}_{n})^{3/5} 
\ {\lambda}^{11/5} \ D^{3/5} \ \ {\rm arcseconds},$$
where $\lambda$ is the wavelength in meters and $D$ is the path length 
through the ISM in kpc.  ${{C}^{2}}\!\!\!{}_{n}$ gives the strength
of electron density fluctuations in the ISM.  Anantharamaiah and  
Narayan \markcite{24} (1988) find ${{C}^{2}}\!\!\!{}_{n} \approx 
1.5$ for low galactic latitudes and $|l| \le {30}^{\circ}$;
we use a value of 1 since the galactic latitude of PKS~1830-211
($-5.\!\!{}^{\circ}7$) is not very small.  Setting $\lambda = 0.013$
meters and $D \approx 5$ kpc (the path length for $|b| = 5.\!\!{}^{\circ}
7$ through a plane parallel ISM with a height of 500 pc) gives us 
${\Theta}_{\rm ISS} \approx 0.4$ mas.  This is also consistent with 
our angular size determination for PKS~1830-211. 

Future work may allow this analysis to be extended to the core of the
NE component.  This will be more difficult because of the 
higher level of extended emission associated with the NE component,
particularly very close to the brightest peak.  If we can determine
the degree of angular broadening of the NE component core due to 
ISS we will have an opportunity to study differences in the 
scattering properties of our galaxy's ISM at multiple frequencies 
along two lines of sight one arcsecond apart. 
The results presented here also suggest that still higher frequency
VLBI observations of PKS~1830-211 will be useful, since the 
intrinsic angular size of the radio source core is still not causing
a significant deviation from the $\theta \propto {\lambda}^2$ 
relation at 22 GHz.  
 
\acknowledgements 

This research was carried out at the
Jet Propulsion Laboratory, 
California Institute of Technology, under contract with the National
Aeronautics and Space Administration.  The Australia Telescope is
operated as a national facility by CSIRO.

\clearpage

\figcaption{
VLBA image of the NE compact component in 
PKS~1830-211 at 22 GHz.  The contours are -1, 1, 2, 4, 8, 16, 32, 50,
70, and 95\%, and the restoring beam (shown in the bottom left corner)
is an elliptical Gaussian with FWHM of $2.2 \times 1.1$ mas and 
the major axis along position angle 1.5$^{\circ}$.
The absolute flux density scale is not well calibrated, but this 
changes only the peak brightness level of the image and not the 
morphology of the source. \label{fig1}} 
\null

\figcaption{
VLBA image of the SW compact component in
PKS~1830-211 at 22 GHz.  The contours are -0.5, 0.5, 1, 2, 4, 8, 16, 32, 
50, 70, and 95\%, and the restoring beam is the same as in figure 1.  
\label{fig2}} 
\null

\figcaption{
The deconvolved minor axis angular size of
the SW component core at 22 GHz ($\lambda = 1.3$ cm), 5 GHz ($\lambda =
6$ cm), 2.3 GHz ($\lambda = 13$ cm), and 1.7 GHz ($\lambda = 18$ cm). 
The deconvolved sizes and their (formal) errors at 22, 5, and 1.7 GHz 
were determined with the AIPS program IMFIT, using only the upper 50\% 
of the brightness range to avoid bias by any extended low-level 
structure.  The size at 2.3 GHz is taken from King (1995).
A least-squares linear fit to the data points is shown; this line
has a slope of $1.96 \pm 0.14$. \label{fig3}} 

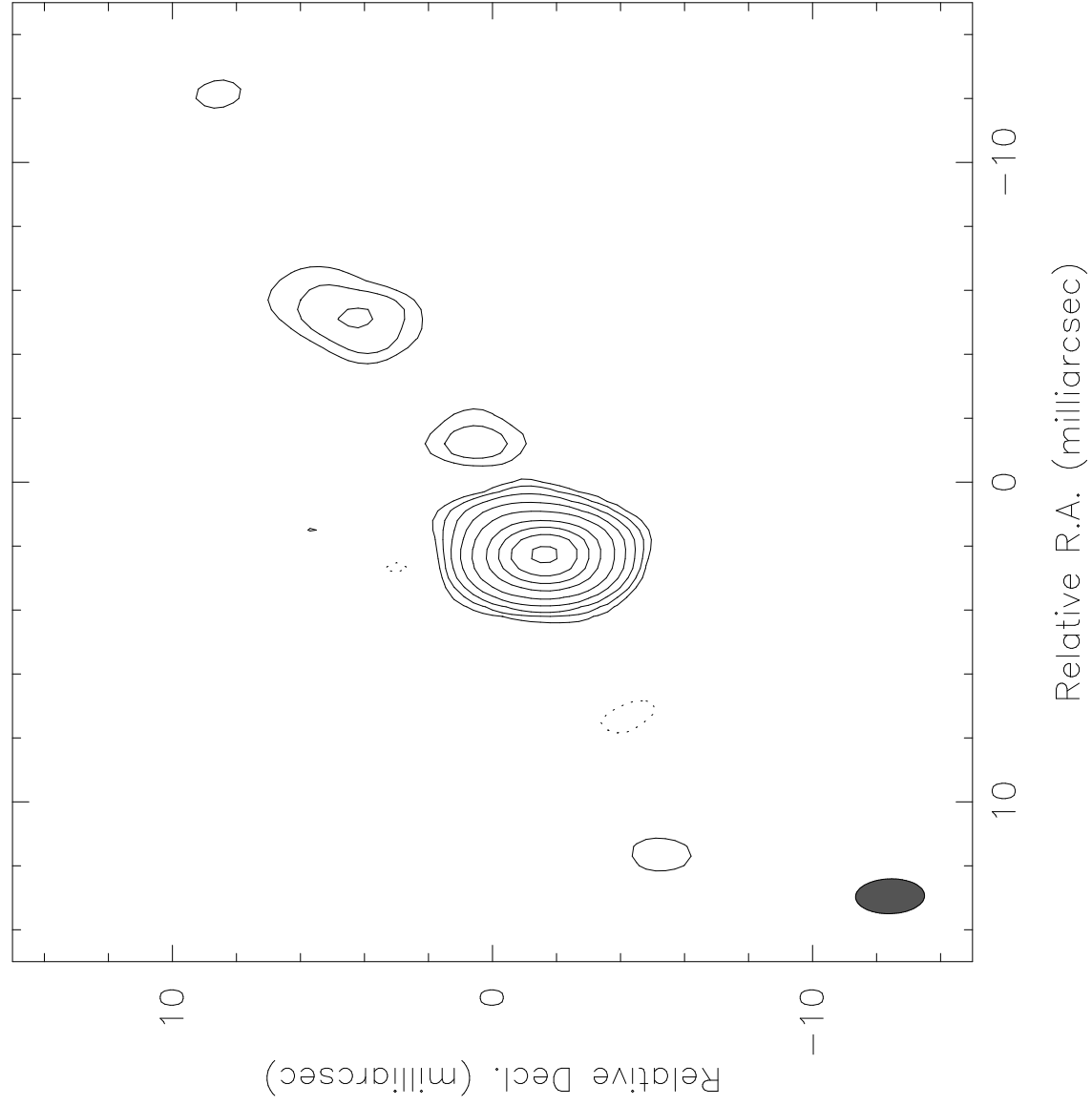
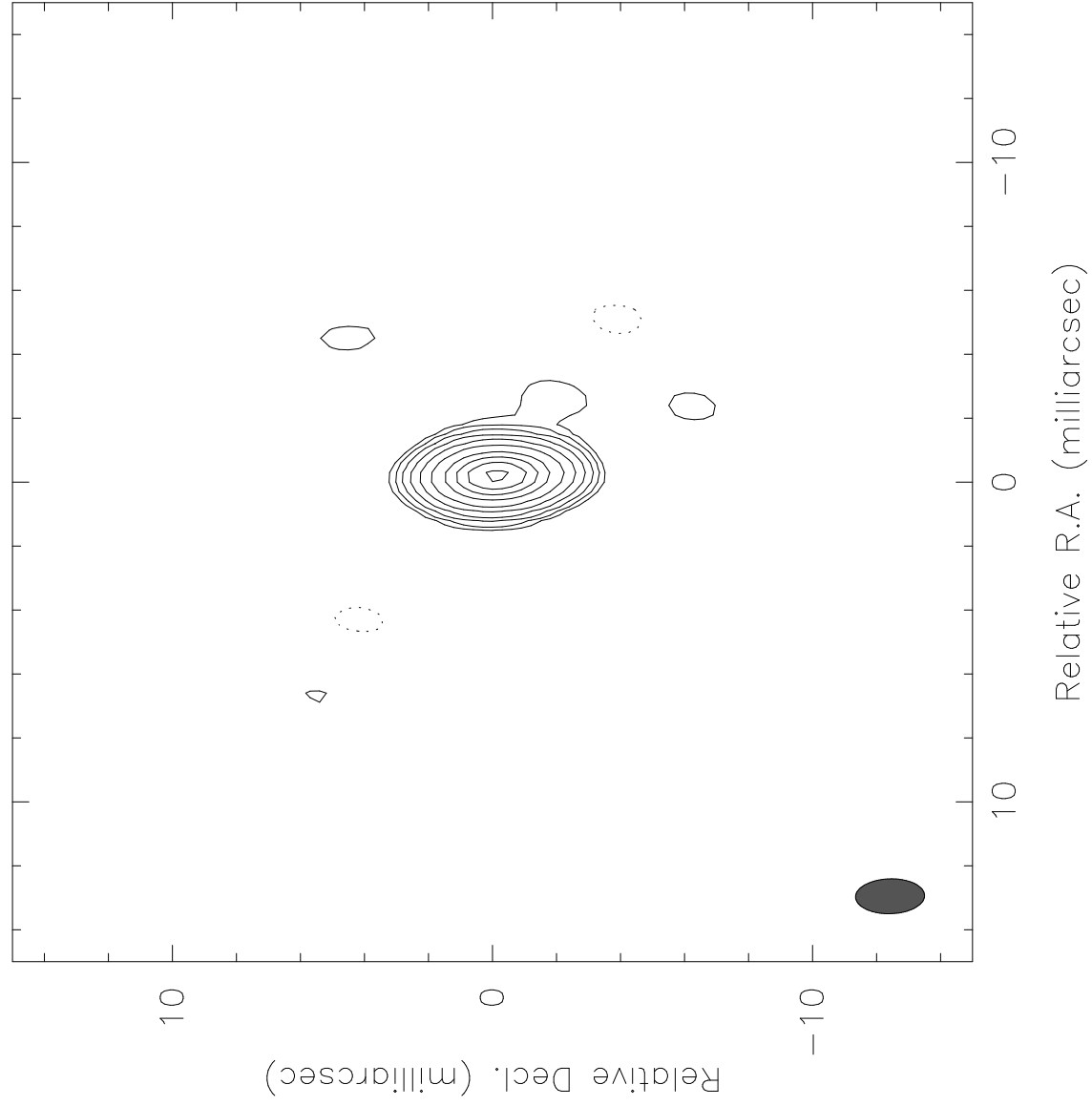
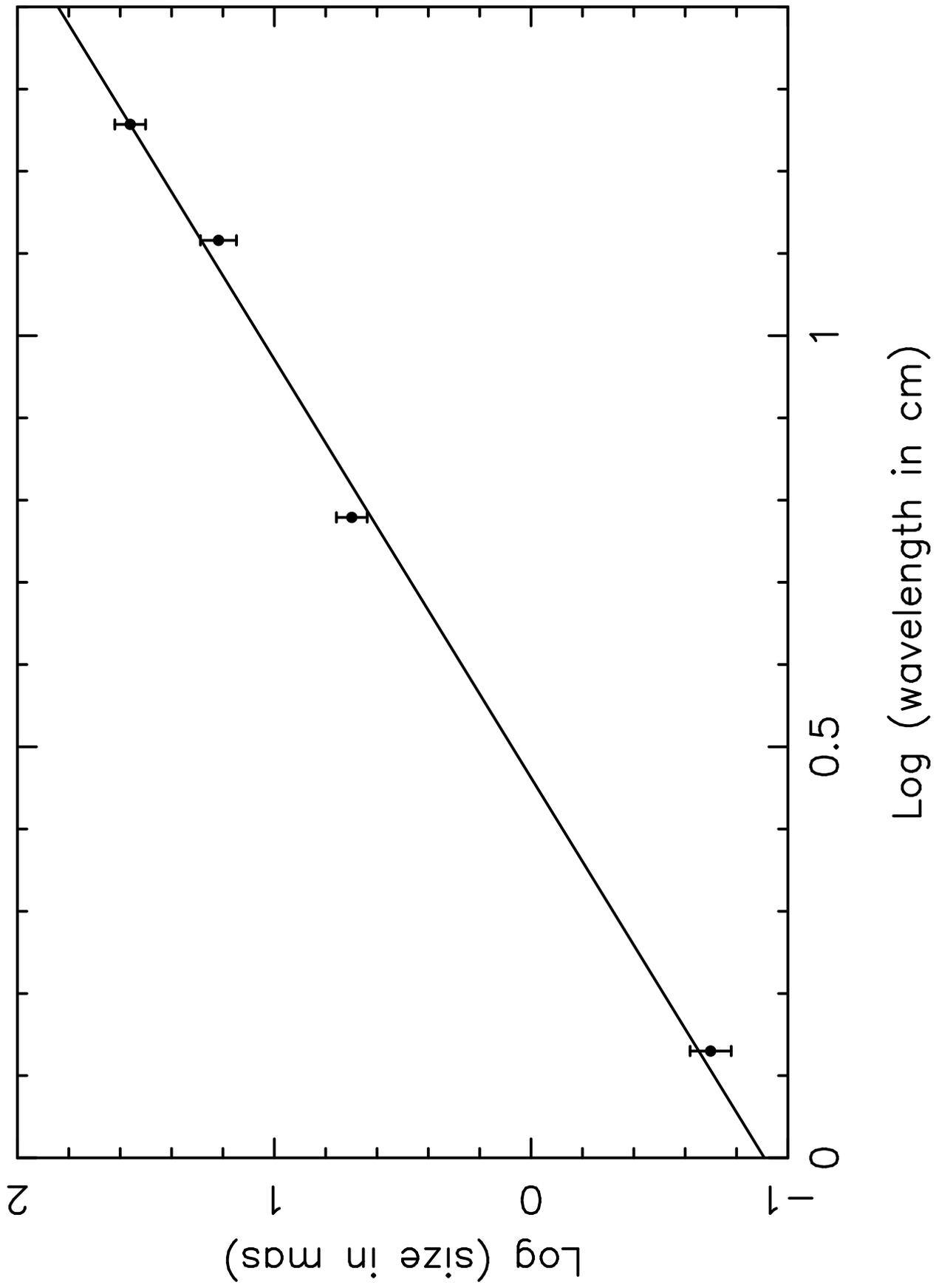
\end{document}